 \documentclass[doublecol,figures]{epl2} 
\usepackage{graphicx}\usepackage{amsmath,amssymb}\usepackage{slashed, bm}
\usepackage{ulem}%

\newcommand{\be}{\begin{equation}}\newcommand{\ee}{\end{equation}}
\newcommand{\bea}{\begin{eqnarray}}\newcommand{\eea}{\end{eqnarray}}

\newcommand{\si}{\sigma}

\newcommand{\Ga}{\varkappa}

\renewcommand{\phi}{\varphi}
\def\Perp{{\scriptscriptstyle \| }}
\newcommand{\p}{p_\Perp}
\def\bp{\mathbf{p}_\Perp}

\def\({\left(}
\def\){\right)}
\def\[{\left[}
\def\]{\right]}
\def\E{{\bf E}}
\def\H{{\bf H}}

\def\S{\bm{ \mathcal S}}
\def\={\mathop{=}}
\def\seq{\mathop{\simeq}}
\renewcommand\Re{\mathop{\rm Re}}
\newcommand{\xx}{{\rm xx}}
\newcommand{\Ref}[1]{(\ref{#1})}

\title{Casimir interaction of strained graphene}

\author{M. Bordag \inst{1} \and I. Fialkovsky\inst{2}\thanks{E-mail: \email{fialkovsky.i@ufabc.edu.br}} \and D. Vassilevich\inst{2,3}}
\shortauthor{M. Bordag, I. Fialkovskiy and D. Vassilevich}

\institute{                    
  \inst{1} Leipzig University, Institute for Theoretical Physics,  04109 Leipzig, Germany \\
  \inst{2} CMCC-Universidade Federal do ABC, Santo Andr\'e, S.P., Brazil\\
	\inst{3} Department of Physics, Tomsk State University, 634050 Tomsk, Russia
}

\abstract{
We calculate the Casimir interaction of two freestanding graphene samples under uniaxial strain. 
Our approach fully takes retardation and dispersion into account and is based on quantum field theoretical expressions for conductivities in termsof the polarization operator.
The force shows a rather weak dependence on the realistic values of strain, changing just by a few percent in its maximum as compared to the non-strained case.}

\begin{document}

\maketitle

\section{Introduction}
The Casimir effect is an important tool for studying new materials
\cite{Woods:2015pla}. Probably the best studied area is the Casimir interaction of graphene. The literature on this topic is already too large to be mentioned in a short paper like the present one, so that we again refer to the review \cite{Woods:2015pla} and to the mini-review \cite{Fialkovsky:2016kio}.

One of the interesting facets of the graphene properties is concerned with its mechanical deformations which called for active research in the recent years, see a comprehensive review \cite{Amorim:2015bga}. In particular, the possibilities of the strain engineering were considered in \cite{Guinea2012}.

The van der Waals and Casimir interactions of the strained graphene layers were also considered.  The van der Waals (non-retarded) interaction was computed in \cite{Sharma2014}. It was found that the force variation is rather small for the strain modulus within the elastic limits, but these variations become strong for an extremely large strain. The Casimir (i.e. fully retarded) interaction of two sheets of strained graphene was first calculated in \cite{Phan:2014} predicting once again quite a strong dependence of the interaction energy on the strain modulus already for moderate values of the strain. However, we have to disagree with these latter results, see Discussions. 

We also like to mention a related calculation of the van der Waals force between an atom and strained graphene \cite{Nichols2016}.

In this letter we report the calculation of the Casimir interaction between free-standing strained graphene which shows that the presence of the strain is hardly of any practical significance giving effect of the order of $6$\% of the already quite small Casimir interaction between graphene samples. 

We shall use the approach based on the polarization tensor and (some) Quantum Field Theory methods 
\cite{Bordag:2009fz,Fialkovsky:2011pu,Bordag:2015zda}. This approach fully takes into account the retardation as well as the momentum dependence of conductivities (dispersion). The main advantage of this approach is that it is consistent with the only experiment on the Casimir interaction of graphene \cite{Banishev:2013}, as has been shown in Ref.\ \cite{Klimchitskaya:2014axa}.

This paper is organized as follows. In the next section we derive the polarization operator for strained graphene basing on the modifications of the graphene microscopical Hamiltonian due to the in-plane strain. In the third section we present the reflection coefficients for a planar anisotropic conducting surface. We conclude the letter with presenting our numerical results for the Casimir interaction energy and discussing its properties, as well as discrepancies with previous research. 

Throughout this paper we shall use the following notations. Latin letters from the beginning of alphabet will correspond to the directions along the surface of graphene, $a,b,c,\ldots =1,2$. The same directions plus the time coordinate will be denoted by indices from the middle of alphabet, $i,j,k,l,\ldots =0,1,2$. We shall use the natural units $\hbar=c=1$.

\section{Polarization tensor and conductivities}
As was demonstrated in \cite{deJuan:2012hxm,DeJuan:2013pha}, a uniform planar strain of the graphene surface leads to the following modified Dirac action for quasiparticles
\begin{equation}
S_D=\int d^3x \, \bar \psi \rho_j^k\gamma^j(i\partial_k-eA_k)\psi \,,\label{Dac}
\end{equation}
where $\gamma^j$ are $2+1$-dimensional $8\times8$ gamma matrices (a direct product of four $2\times2$ irreducable ones), $A_l$ is the electromagnetic potential, and
\begin{equation}
\rho_0^0=1\,,\quad \rho_0^a=\rho_a^0=0\,,\quad \rho_a^b=v_{ab}\,.\label{rho}
\end{equation}
The matrix $v_{ab}$ can be interpreted as a tensorial Fermi velocity,
see \cite{DeJuan:2013pha}
\begin{equation}
	v_{ab}= v_F \left[ \delta_{ab}-\frac\beta4\( 2u_{ab}+ \delta_{ab} u_{cc}\)\right] \label{vab}.
\end{equation}
that replaces the usual scalar Fermi velocity $v_F\simeq 1/300$. For graphene,  $\beta\simeq 2$.
$u_{ab}$ is the strain tensor. Taking the direction of uniaxial uniform planar strain to have the angle $\theta$ with the $x$-direction, one gets \cite{Pereira:2009}
\be
	u=\epsilon 
	\(\begin{array}{cc}
		\cos^2\theta-\sigma \sin^2\theta & (1+\sigma) \sin\theta \cos\theta\\
		(1+\sigma) \sin\theta \cos\theta & \sin^2\theta-\sigma \cos^2\theta
	\end{array}\).
\ee
Here $\epsilon$ is the strain modulus that may reach $0.2$ for elastic deformations. $\sigma$ is the Poisson ratio. Possible values of $\sigma$ are discussed in \cite{Farjam:2009}. We take $\sigma=0.14$. One can easily check, that conductivities and thus the Casimir energy are not sensitive to small variations of $\sigma$.

At zero temperature, the polarization tensor for the Dirac theory (\ref{Dac}) can be written as
\begin{eqnarray}
&&\tilde \Pi^{jk}(p)=ie^2\int \frac {d^3k}{(2\pi)^3}\, {\rm tr} \left( \rho^j_l\gamma^l \hat S(k)
\rho^k_m\gamma^m \hat S(k-p) \right)\nonumber\\
&&\quad = ie^2\rho^j_l\rho^k_m\int \frac {d^3k}{(2\pi)^3}\, {\rm tr} \left( \gamma^l \hat S_0(\rho k)
\gamma^m \hat S_0(\rho(k-p)) \right)\nonumber
\end{eqnarray}
where $\hat S$ is the Greens' function of the Dirac operator $\slashed{D}=i\rho^j_k\gamma^k\partial_j$, while $\hat S_0$ is the Greens' function of the same operator without rescaling $\slashed{D}_0=i\gamma^j\partial_j$. By making the change of integration variables $k\to \rho k$ one arrives at the identity
\begin{equation}
  \tilde \Pi^{jk}(p) =(\det \rho)^{-1}\rho^j_l\rho^k_m  \Pi^{lm}(\rho p)\,. \label{PiPi}
\end{equation}
This formula relates the polarization tensor $\tilde\Pi$ for the strained graphene to another polarization tensor, $\Pi$, which is calculated without strain and for the unit Fermi velocity. One can easily prove that the relation (\ref{PiPi}) holds (and has the form exactly as written above) also for a non-zero temperature and in the presence of a mass gap and of a chemical potential.

The expressions for $\Pi^{ij}$ can be found in Refs.\ \cite{Bordag:2009fz,Fialkovsky:2011pu,Bordag:2015zda}. One should remember to put $v_F=1$ is that expressions. Here we present a short summary.

Due to the symmetry properties, the parity-even part of the polarization tensor depends on two functions $A$ and $B$ of the momenta
\be
\Pi^{ji}= 
    \Pi^{ji}_0 A(p_0,p_a)
    +p_0^2 \Pi^{ji}_u  B(p_0,p_a)\,,
\ee
where
\begin{eqnarray}
&&\Pi^{ji}_0
    =g^{ji}-\frac{  p^j  p^i}{  p^2},\\
&&\Pi^{ji}_u
    =\frac{  p^j  p^i}{  p^2}-\frac{  p^j u^i + u^j   p^i}{(  pu)}
    +\frac{u^ju^i}{(  pu)^2}  p^2 . \nonumber
\end{eqnarray}
In the medium rest reference frame $u=(1,0,0)$. It is convenient to use other two independent functions,
\begin{equation}
\Pi_{00}\quad \mbox{and}\quad \Pi_{\rm str}=\Pi_{11}+\Pi_{22}\,.
\end{equation}
The polarization tensor may be separated into a ``vacuum" part $\Pi^{(vac)}$ corresponding to $T=\mu=0$ and the rest, denoted by $\Delta\Pi$:
\be 
  \Pi_\xx(p;\mu, T) 	= \Pi_\xx^{(vac)}(p) 		+\Delta \Pi_\xx (p;\mu,T)\,,
  \label{decomp1} 
\ee
where $\xx$ stands  for either `${\rm str}$' or `$00$'.

The Casimir energy is defined by the Lifshitz formula (see eq.\ (\ref{EL}) below) as an integral and a sum over Euclidean momenta. Therefore, we perform now the Wick rotation $p_0\to p_4=ip_0$ and stay in Euclidean momentum space till the end of this section. Let us define 
\begin{equation}
\bp=(p_1,p_2)\,\qquad\p=|\bp|\,,\qquad p^2=p_4^2+\p^2\,.\label{ppp}
\end{equation}
Then,
\begin{eqnarray}
&&\Pi^{\rm (vac)}_{00}(p)=\frac{\alpha \p^2 \Phi (p)}{ p^2} \,,\nonumber\\
&&\Pi^{\rm (vac)}_{\rm str}(p)=-\frac{\alpha   (2p_4^2+ \p^2)\Phi(p)}{p^2}
\,,\label{Pivac}
\end{eqnarray}
where
\begin{equation}
\Phi=4 \left[ m + \frac{  p^2 -4m^2}{  2 p} 
\arctan \left(\frac{ {p}}{2m}\right)\right].
\end{equation}
For the parts of the polarization tensor that depend on $\mu$ and $T$, we have \cite{Bordag:2015zda}
\begin{eqnarray}
&&\Delta\Pi_{00} (p) =\label{DP00}\\
&&={ 8\alpha  }
	\int_m^\infty \!d\varkappa
	\(1+ \Re\frac{- p^2+4i p_4 \Ga+4\Ga^2}{\sqrt{Q^2-4\p^2 (\Ga^2-m^2)}} \) \Xi (\varkappa)
	\nonumber
\end{eqnarray}
and
\begin{eqnarray}
&&\Delta\Pi_{\rm str} (p) =\label{DPstr}\\
&&={ 8\alpha   }
	\int_m^\infty \!d\varkappa 
	 \Re\frac{\(4i p_4 \Ga+4\Ga^2 -4m^2\)\Xi (\varkappa) }{\sqrt{Q^2-4\p^2 (\Ga^2-m^2)}} ,
	\nonumber
\end{eqnarray}
we have used the notation $Q= p^2 -2i p_4 \Ga$,
\begin{equation}
\Xi\equiv {(e^{(\Ga+\mu)/T}+1)^{-1}} + {(e^{(\Ga-\mu)/T}+1)^{-1}} 
  \label{Xi}
\end{equation}
is the distribution function. $\alpha$ is the fine structure constant. With our conventions 
$\alpha=\tfrac{e^2}{4\pi}=\tfrac 1{137}$.

Note that \Ref{Pivac}--\Ref{DPstr} for the polarization tensor take into account all species of quasiparticles in graphene which consist of $N=4$ copies of a $2$-component fermion.

Other calculations of the optical conductivity of graphene were performed in
\cite{OlivaLeyva:2014} using, however, a different strain model and disregarding the spatial dispersion. The optical conductivity of graphene under some other geometrical distortions of the crystal lattice was considered in \cite{Sinner:2010vc} (random lattice deformations) and in \cite{Chaves:2013fca} (out-of-plane deformations).

\section{Reflection coefficients} Here we derive the reflection coefficients on the surface of graphene without specifying any particular form for the polarization tensor. For simplicity, we suppose that graphene is suspended in vacuum. Let the surface of graphene occupy the plane at the constant value of the coordinate $x^3=0$. The matching conditions for the electromagnetic potential $A_\mu$ read \cite{Bordag:2009fz}
\begin{equation}
[A_\mu]=0\,,\qquad [\partial_3 A_\mu]=\tilde \Pi_\mu^{\ \nu}A_\nu\,,\label{matchA}
\end{equation}
where the square brackets denote the discontinuity of corresponding function on the surface of graphene. $\mu,\nu=0,1,2,3$ are space-time indices. By definition, $\tilde\Pi^{\mu 3}=\tilde\Pi^{3\mu}=0$. Due to the gauge invariance, the polarization tensor has to satisfy the transversality conditions $p_j\tilde \Pi^{jk}=0$ with $p_j$ being the momentum of electromagnetic field. Due to these conditions, one may express the temporal components of $\tilde \Pi$ as
\begin{equation}
\tilde \Pi^{0a}=-\frac{p_b\tilde\Pi^{ba}}{p_0}\,,\qquad \tilde\Pi^{00}=\frac{p_a\tilde\Pi^{ab}p_b}{p_0^2}\,, \label{Piab}
\end{equation}
which together with the definition of conductivities $\sigma_{ab}$ \cite{Fialkovsky:2016kio}
\begin{equation}
\sigma_{ab}=\frac{\tilde \Pi^{ab}}{ip_0}\label{sigab}
\end{equation}
permits to rewrite eventually the matching conditions (\ref{matchA}) through the electric and magnetic fields and the conductivities:
\begin{eqnarray}
&&[E_a]=[H_3]=0\,,\label{mEa}\\
&&[E_3]=-\frac{i p_aE_b\tilde\Pi^{ab}}{p_0^2},\nonumber\\
&&\quad	=\frac{p_1 E_1 \si_{11}
		+(p_2 E_1 +p_1 E_2)\si_{12}
			+ p_2 E_2 \si_{22}}{p_0},\label{mE3}\\
&&[H_1]=-\frac{i\tilde\Pi^{a2}E_a}{p_0}=\si_{12}E_1+\si_{22}E_2,\label{mH1}\\
&&[H_2]=\frac{i\tilde\Pi^{a1}E_a}{p_0}=-\si_{11}E_1-\si_{12}E_2.\label{mH2}
\end{eqnarray}
We remark, that to derive these equations the gauge condition $A_3=0$ was useful. We have also assumed that the polarization tensor, and thus the conductivity, is symmetric, $\sigma_{12}=\sigma_{21}$.

Next, we have to define the scattering matrix. Our approach is most close to that of Ref.\ \cite{Tse:2011}.  The scattering field is defined as
\be
	\E=\left\{
		\begin{array}{ll}
		\E^-_i e^{ip_3 x^3}
			+\E^-_r e^{-ip_3 x^3} &x^3<0\\
		\E^+_i e^{-ip_3 x^3}
			+\E^+_r e^{ip_3 x^3} &x^3>0
		\end{array}
		\right.
\ee
where index $i,r$ stands for incident or reflected waves, respectively, and the vector components are 
\be
	\E^\pm_{i,r}= e^{i p_a x^a} \(e^{\pm i,r}_1,e^{\pm i,r}_2,e^{\pm i,r}_3\)^T.
\ee
The magnetic field is defined through the Maxwell equations,
\be
  \H=\frac{-i}{p_0} \nabla \times \E ,\qquad 
    \nabla \equiv (\partial_1,\partial_2,\partial_3).
\ee
The scattering problem is solved by a $4\times 4$ matrix $\S$ in the following sense \cite{Tse:2011}
\be
	\(e^{- r}_1,e^{-r}_2,e^{+ r}_1,e^{+r}_2\)^T = 
		\S	\(e^{- i}_1,e^{-i}_2,e^{+ i}_1,e^{+i}_2\)^T
\ee 
(The third component $e^{\pm i,r}_3$ is a dependent variable and can be excluded).
This matrix can be further separated in $2\times 2$ blocks:
\be
\S=		\(\begin{array}{ll}
		\mathbf{R} & \mathbf{T}\\
		\mathbf{T} & \mathbf{R}
		\end{array}
		\),\quad
	\mathbf{T} = {\bm 1} + \mathbf{R}
\ee
After long but straightforward calculations we arrive at the following result
\begin{equation}
	\mathbf{R}= \frac 1{\Delta}
	\(\begin{array}{cc}
		r_{xx} & r_{xy}\\
		r_{yx}& r_{yy}
		\end{array}
		\),
		\label{R}
\end{equation}
where
\begin{eqnarray}
&&r_{xx}=\sigma_{12}^2-\sigma_{11}\sigma_{22} +\frac{2\bigl( p_1p_2\sigma_{12}-
(p_2^2+p_3^2)\sigma_{11} \bigr)}{p_0p_3}\,,\nonumber\\
&&r_{yy}=\sigma_{12}^2-\sigma_{11}\sigma_{22} +\frac{2\bigl( p_1p_2\sigma_{12}-
(p_1^2+p_3^2)\sigma_{22} \bigr)}{p_0p_3}\,,\nonumber\\
&&r_{xy}=\frac 2{p_0p_3} \bigl( p_1p_2\sigma_{22} -(p_3^2+p_2^2),\sigma_{12} \bigr)\nonumber\\
&&r_{yx}=\frac 2{p_0p_3} \bigl( p_1p_2\sigma_{11} -(p_3^2+p_1^2)\sigma_{12} \bigr)
\end{eqnarray}
and
\begin{eqnarray}
&&\Delta=\frac 2{p_0p_3} \bigl(\sigma_{11}(p_3^2+p_2^2)+\sigma_{22}(p_3^2+p_1^2)-2p_1p_2\sigma_{12}
\bigr) \nonumber\\
&& \qquad + 4 + \sigma_{11}\sigma_{22} -\sigma_{12}^2,
\end{eqnarray}
while $p_3=+\sqrt{p_0^2-p_1^2-p_2^2}$.

We note here that the obtained expression for $\mathbf{R} $ is different from the  Eq.\ (2) of  \cite{Phan:2014} taken at $\varepsilon=\varepsilon_0=\mu_0=1$, see the discussion in the last section.

\section{Casimir energy} The Casimir free energy of two surfaces characterized by their reflection matrices ${\mathbf R},$ and separated by distance $a$, is defined by the Lifshitz formula \cite{Lifshitz}

\be
    {\mathcal F}
    =T\sum_{n=-\infty}^\infty\int\frac{d^2{\bf p}}{8\pi^2} 
      \ln \det \left[ 1-  {\mathbf R}_1(p) {\mathbf R}_2(p) \, e^{-2p a}\right] \,,
        \label{EL}
\ee
which uses the Euclidean momenta. The polarization tensor \Ref{decomp1}-\Ref{Xi} has already been written in this signature. One should not forget to Wick-rotate the reflection matrix ${\mathbf R},$  Eq. (\ref{R}). The notations (\ref{ppp}) have been used. In addition, $p_4=2\pi nT$, where $T$ is the temperature. For the case of zero temperature, one just needs to substitute the sum with an integral, $T\sum_n\to\int \frac{dp_4}{2 \pi } $. 

\begin{figure}
\onefigure[width=8cm]{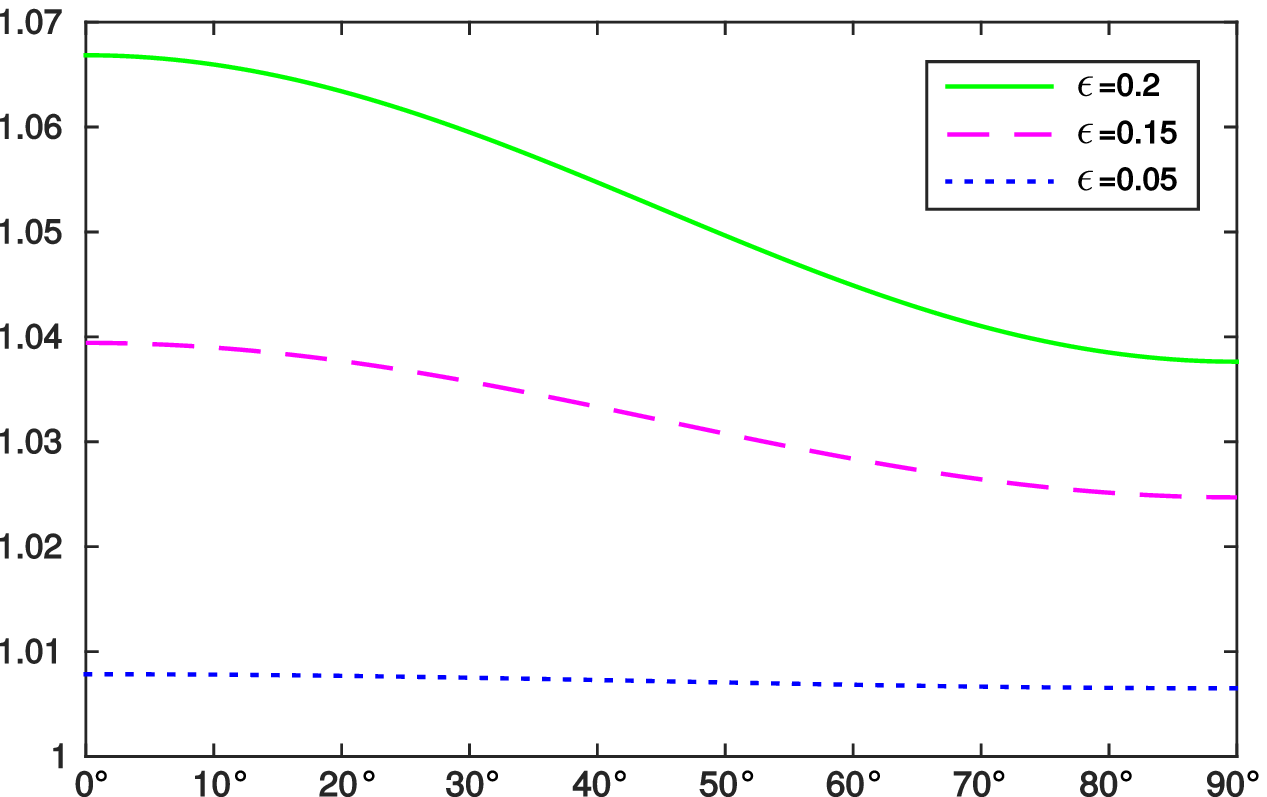}
\caption{Angular dependence of the Casimir energy for $T=0$, $a=30{\rm nm}$ and different values of $\epsilon$, normalized to the unstrained case with the same temperature and separation.}
\label{fig1}
\end{figure}

\begin{figure}
\onefigure[width=8cm]{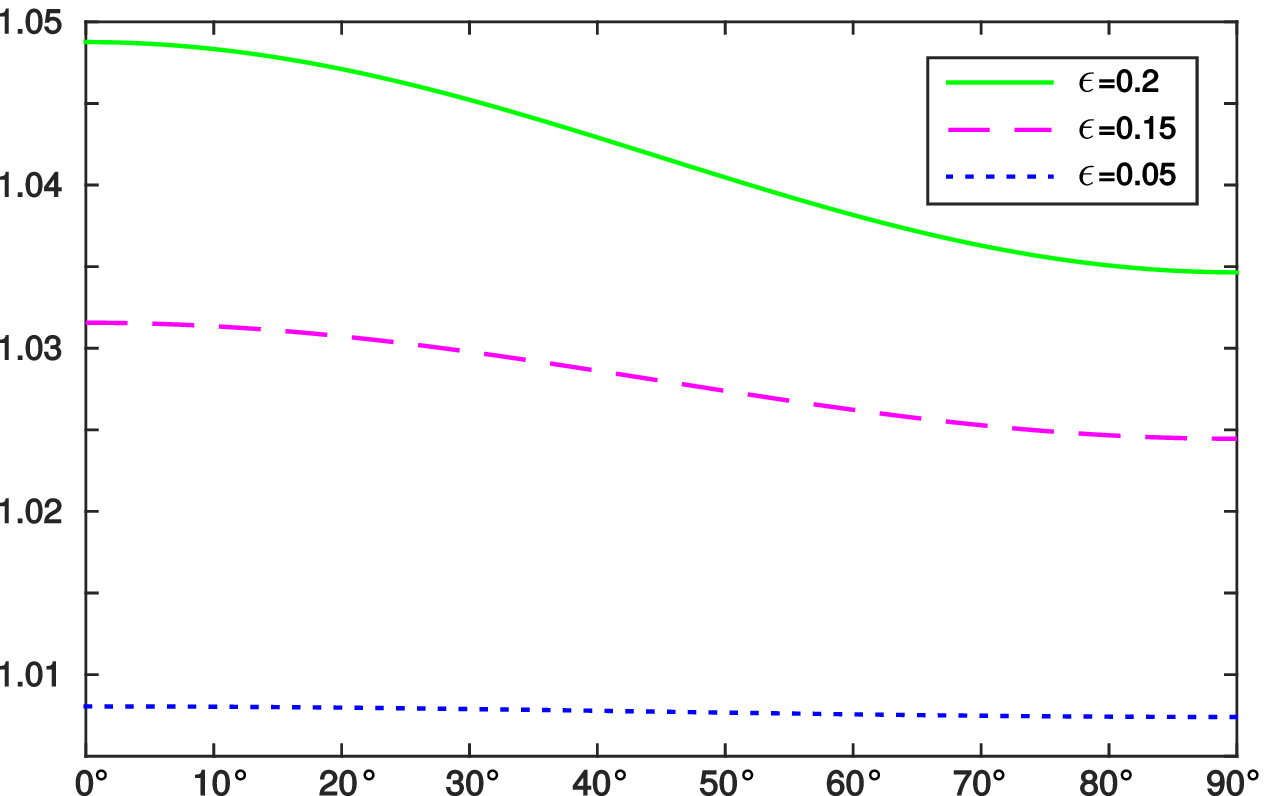}
\caption{Same as Fig.\ref{fig1} but for $T=300K$.}
\label{fig2}
\end{figure}

\begin{figure}
\onefigure[width=8cm]{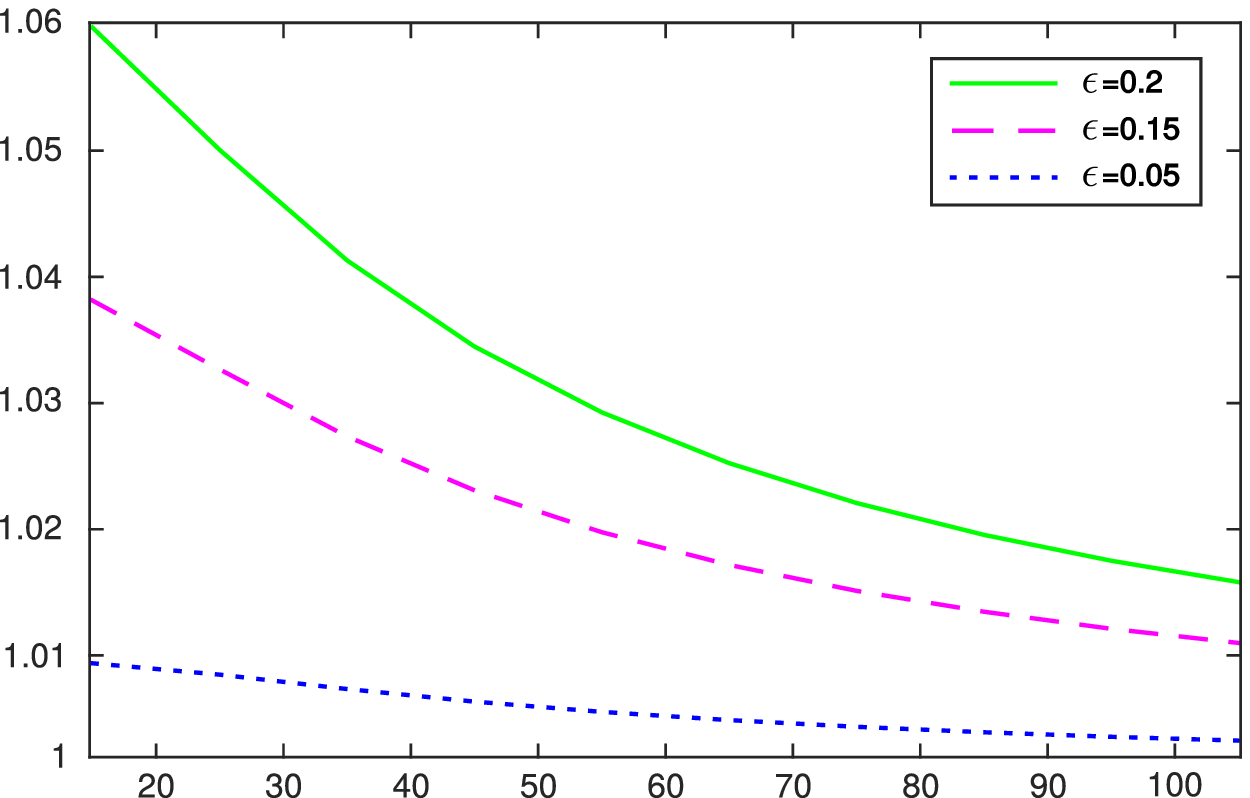}
\caption{Distance dependence of the Casimir energy for $T=300K$ and the angle $\theta=\pi/6$, normalized to the unstrained case with the same temperature.}
\label{fig4}
\end{figure}

Before proceeding with strained graphene layers, it is instructive to set a reference point and  reproduce the Casimir interaction between unstrained pristine layers. Evaluation of the ratio of \Ref{EL} at $\epsilon =0$ to the Casimir energy of two ideal metals, ${\mathcal F}_0=-\pi^2/720 a^3$, both at zero temperature, shows
\be
  \frac{{\mathcal F}({\epsilon =0})}{{\mathcal F}_0}\seq 0.00485.
\ee
This value is close to but slightly lower then those obtained via the models of constant conductivity\cite{Drosdoff2010,Khusnutdinov2014} or in non-retarded calculations \cite{Sernelius2011}, and coincides with results of \cite{Klimchitskaya2013}.

Turning now to the main subject of our paper, we assume that the strain modulae on both graphene samples are equal. On one of the layers strain is directed along $x^1$, while in the second one its direction forms an angle $\theta$ with $x^1$. Then the whole Casimir energy \Ref{EL} becomes a function of both the distance $a$, and of the angle $\theta$, $ {\mathcal F}= {\mathcal F}(a,\theta)$. The results of the numerical simulations of the latter are depicted at Fig. \ref{fig1}-\ref{fig4}.

We restrict ourselves to the values of strain modulus $\epsilon\leq0.2$. Larger values cause non-elasticity of the deformation and various types of instabilities \cite{Liu:2007,Lee:2008,Si:2016}.

 The analysis of Fig. \ref{fig1}-\ref{fig4} representing the results of our numerical simulation reveals that for the values of strain modulus $\epsilon\leq0.2$ the variation the Casimir interaction between two strained (and otherwise free-standing) graphene samples does not reach more then $6.7\%$ at its maximum at zero temperature and separation of $30$nm.  At the room temperature, $T=300K$, the effect diminishes, but not that much, to about $5\%$. 
In all calculations we considered $m=\mu=0$, though for the purpose of numerical evaluation we gave them a very small value of $10^{-8}$eV.

At larger separations the influence of strain is even less.  The distance dependence of the strained Casimir energy is presented at  Fig. \ref{fig4}. As one can notice, the presence of the strain does change it, however only subtly at small separations, and vanishingly small at $100$nm, or higher. 

We  do not observe any angle where the Casimir energy would be independent of the strain modulus in the interacting samples, contrary to observations of \cite{Phan:2014}. 

\section{Discussion}
First of all, let us estimate very roughly the effect of strain on the Casimir interaction of graphene layers. To this end, let us adopt the constant conductivity model and neglect any dispersion.
Then, the Casimir interaction is proportional to the conductivity\cite{Khusnutdinov2014}. If the conductivity of unstrained graphene is diagonal with $\sigma_{11}=\sigma_{22}:=\sigma_0$, under the strain in the $x^1$-direction with the modulus $\epsilon=0.2$ it changes as $\sigma_{11}\simeq 0.7\sigma_0$, $\sigma_{22}\simeq \sigma_0/0.7$. Consider the case of parallel strains in both samples. If we neglect any mixing between $x^1$- and $x^2$-polarizations, the contribution to the energy of the former is multiplied by a factor of $0.7$, while of the latter --- divided by the same factor. The overall effect is thus 
an enhancement of the Casimir energy by about $6.4\%$.  This naive estimate is in a very good agreement with our exact calculations at $T=0$, see Fig.~\ref{fig1}.

A detailed comparison between our results and those of non-retarded calculations of \cite{Sharma2014} is hardly possible. The models are too different and there is no clear way to identify the parameters. However, at a qualitative level there is no disagreement. The paper \cite{Sharma2014} predicted a rather small effect for moderate strain modulus that is maximal for parallel strains in both samples, which is in an agreement with our findings. The anisotropy for which the enhancement of the van der Waals interaction of \cite{Sharma2014} becomes strong corresponds to the strain values of approximately $\epsilon=0.8$. We do not consider so high values of the strain modulus, for the reason that has been already explained above.

Although our methods are closer to those of the paper \cite{Phan:2014}, we disagree with the findings in that work. In particular, Ref. \cite{Phan:2014} claims very large variation of the Casimir energy already for moderate values of strain. The main difference in our approaches is that we took the dispersion into account, while \cite{Phan:2014} used a constant conductivity model. We also disagree with the expression for reflection coefficients in \cite{Phan:2014}, but cannot reliably trace back the source of this difference.

Finally, a comment on the strain model is in place. There is an expression for $v_{ab}$ in the literature \cite{OlivaLeyva:2013,OlivaLeyva:2014} that is different from (\ref{vab}). We use Eq. (\ref{vab}) since we find the arguments of \cite{deJuan:2012hxm,DeJuan:2013pha} more convincing. Though the use of the model \cite{OlivaLeyva:2013,OlivaLeyva:2014} would change the numerical values of the Casimir energy, it could not affect the overall smallness of the effect.

\acknowledgments
This work was supported in part by the grant 2016/03319-6 of S\~ao Paulo Research Foundation (FAPESP),  by the grant 303807/2016-4 of CNPq, and by the Tomsk State University Competitiveness Improvement Program.

\bibliography{graphene}

\begin{thebibliography}{10}
\expandafter\ifx\csname url\endcsname\relax\def\url#1{\texttt{#1}}\fi

\bibitem{Woods:2015pla}
\Name{Woods L., Dalvit D., Tkatchenko A., Rodriguez-Lopez P., Rodriguez A. \and
  Podgornik R.} \REVIEW{Rev. Mod. Phys.}{88}{2016}{045003}.

\bibitem{Fialkovsky:2016kio}
\Name{Fialkovsky I.~V. \and Vassilevich D.~V.} \REVIEW{Mod. Phys. Lett.A}{
  31}{2016}{1630047}.

\bibitem{Amorim:2015bga}
\Name{Amorim B. \etal} \REVIEW{Phys. Rept.}{617}{2016}{1}.

\bibitem{Guinea2012}
\Name{Guinea F.} \REVIEW{Solid State Communications}{152}{2012}{1437}.

\bibitem{Sharma2014}
\Name{Sharma A., Harnish P., Sylvester A., Kotov V.~N. \and Castro~Neto A.~H.}
  \REVIEW{Phys. Rev.B}{ 89}{2014}{235425}.

\bibitem{Phan:2014}
\Name{Phan A.~D. \and Phan T.} \REVIEW{Phys. Status Solidi RRL}{8}{2014}{1003}.

\bibitem{Nichols2016}
\Name{Nichols N.~S., Maestro A.~D., Wexler C. \and Kotov V.~N.} \REVIEW{Phys.
  Rev.B}{ 93}{2016}{205412}.

\bibitem{Bordag:2009fz}
\Name{Bordag M., Fialkovsky I.~V., Gitman D.~M. \and Vassilevich D.~V.}
  \REVIEW{Phys. Rev.B}{ 80}{2009}{245406}.

\bibitem{Fialkovsky:2011pu}
\Name{Fialkovsky I.~V., Marachevsky V.~N. \and Vassilevich D.~V.} \REVIEW{Phys.
  Rev.B}{ 84}{2011}{035446}.

\bibitem{Bordag:2015zda}
\Name{Bordag M., Fialkovskiy I. \and Vassilevich D.} \REVIEW{Phys. Rev.B}{
  93}{2016}{075414}.

\bibitem{Banishev:2013}
\Name{Banishev A.~A., Wen H., Kawakami R.~K., Klimchitskaya G.~L., Mostepanenko
  V.~M. \and Mohideen U.} \REVIEW{Phys. Rev.B}{ 87}{2013}{205433}.

\bibitem{Klimchitskaya:2014axa}
\Name{Klimchitskaya G.~L., Mohideen U. \and Mostepanenko V.~M.} \REVIEW{Phys.
  Rev.B}{ 89}{2014}{115419}.

\bibitem{deJuan:2012hxm}
\Name{de~Juan F., Sturla M. \and Vozmediano M. A.~H.} \REVIEW{Phys. Rev.
  Lett.}{108}{2012}{227205}.

\bibitem{DeJuan:2013pha}
\Name{De~Juan F., Ma\~nes J.~L. \and Vozmediano M. A.~H.} \REVIEW{Phys. Rev.B}{
  87}{2013}{165131}.

\bibitem{Pereira:2009}
\Name{Pereira V.~M., Castro~Neto A.~H. \and Peres N. M.~R.} \REVIEW{Phys.
  Rev.B}{ 80}{2009}{045401}.

\bibitem{Farjam:2009}
\Name{Farjam M. \and Rafii-Tabar H.} \REVIEW{Phys. Rev.B}{80}{2009}{167401}.

\bibitem{OlivaLeyva:2014}
\Name{Oliva-Leyva M. \and Naumis G.~G.} \REVIEW{J. Phys: Condens.
  Matter}{26}{2014}{125302} errata - ibid. 26 (2014) 279501.

\bibitem{Sinner:2010vc}
\Name{Sinner A., Sedrakyan A. \and Ziegler K.} \REVIEW{Phys. Rev.B}{
  83}{2011}{155115}.

\bibitem{Chaves:2013fca}
\Name{Chaves A.~J., Frederico T., Oliveira O., de~Paula W. \and Santos M.~C.}
  \REVIEW{J. Phys. Condens. Matter}{26}{2014}{185301}.

\bibitem{Tse:2011}
\Name{Tse W.-K. \and MacDonald A.~H.} \REVIEW{Phys. Rev.B}{ 84}{2011}{205327}.

\bibitem{Lifshitz}
\Name{Lifshitz E.~M. \and Pitaevskii L.~P.} \Book{{Statistical Physics: Part
  2}} (Pergamon Press, Oxford, UK) 1980.

\bibitem{Drosdoff2010}
\Name{Drosdoff D. \and Woods L.~M.} \REVIEW{Phys. Rev.}{82}{2010}{155459}.

\bibitem{Khusnutdinov2014}
\Name{Khusnutdinov N., Drosdoff D. \and Woods L.~M.} \REVIEW{Phys. Rev.D}{
  89}{2014}{085033}.

\bibitem{Sernelius2011}
\Name{Sernelius B.~E.} \REVIEW{EPL (Europhysics Letters)}{95}{2011}{57003}.

\bibitem{Klimchitskaya2013}
\Name{Klimchitskaya G. \and Mostepanenko V.} \REVIEW{Phys. Rev.
  B}{87}{2013}{075439}.

\bibitem{Liu:2007}
\Name{Liu F., Ming P. \and Li J.} \REVIEW{Phys. Rev.B}{ 76}{2007}{064120}.

\bibitem{Lee:2008}
\Name{Lee C., Wei X., Kysar J.~W. \and Hone J.}
  \REVIEW{Science}{321}{2008}{385}.

\bibitem{Si:2016}
\Name{Si C., Sun Z. \and Liu F.} \REVIEW{Nanoscale}{8}{2016}{3207}.

\bibitem{OlivaLeyva:2013}
\Name{Oliva-Leyva M. \and Naumis G.~G.} \REVIEW{Phys. Rev.B}{
  88}{2013}{085430}.

\end{thebibliography}
\bibliographystyle{eplbib}

\end{document}